\documentstyle[preprint,prd,aps,psfig,floats]{revtex}
\newcommand{\be}{\begin{equation}}
\newcommand{\ee}{\end{equation}}
\newcommand{\bea}{\begin{eqnarray}}
\newcommand{\eea}{\end{eqnarray}}

\tightenlines
\begin{document}
\draft
%
\firstfigfalse
%
\title{Bottom and Charm Production at LHC and RHIC}
\author{Jens Ole Schmitt, Gouranga C. Nayak, 
Horst St\"ocker and Walter Greiner}
\address
{\small\it{Institut f\"ur Theoretische Physik,
J. W. Goethe-Universit\"at,
60054 Frankfurt am Main, Germany}}
\maketitle

\begin{abstract} 
We study $b\bar{b}$ and $c\bar{c}$ production and the influence of nuclear shadowing 
at LHC and RHIC energies. 
We find a significant reduction 
in the production cross section of both charm and bottom at RHIC and LHC.
Bound states
such as $\Upsilon$ and $J/\Psi$ are suppressed by this reduction 
in the charm production
cross sections. Therefore, $J/\Psi$ suppression may not be useful 
as a signature for the quark gluon plasma. 
\end{abstract}  
\bigskip

%
As lattice QCD results indicated, hadronic matter probably undergoes a phase
transition into a deconfined quark-gluon plasma (QGP) phase, provided a
temperature about 200 MeV or an energy density $\sim$ 2
GeV/fm$^3$ is reached \cite{lattice}. High energy heavy-ion colliders at 
RHIC (Au-Au at $\sqrt s$=200 GeV) and LHC (Pb-Pb at $\sqrt s$=5.5 TeV) will
provide the best opportunity to study the formation and evolution of 
quark-gluon plasma \cite{torino}. As QGP exists only for a very short time 
(several fermi) in a small volume ($\sim$ 100 fm$^3$), 
a direct detection of this state of matter is not possible. Thus various
indirect signatures have to be used for its detection.
The prominent ones among the suggested signatures are $J/\Psi$ suppression 
\cite{satzs}, strangeness enhancement \cite{rafelski,singh}, 
dilepton and photon production \cite{strickland,shuryak,alam,nayak-dilep}.
Although these signatures are proposed, it is not easy to make a concrete 
calculation of them for these experiments. Many uncertainties, 
both theoretical and experimental, exist which make it very difficult
to claim the existence of the quark-gluon plasma based on the measurement
of these signatures. At the present experiment
at SPS most of the signatures can be explained
by alternative methods without assuming the existence of quark-gluon plasma.
In this paper we will study the impact of such non-QGP sources
on the production of $J/{\Psi}$ at RHIC and LHC energies. 

We mention here that due to the presence of nuclear, quark-gluon
and hadronic medium a correct computation of $J/\Psi$ 
suppression in ultra relativistic 
heavy-ion collisions (URHIC) becomes extremely difficult.  
We highlight here the major uncertainties associated with the
calculation of $J/{\Psi}$ suppression. First of all one
has to understand the production of $J/{\Psi}$ in nuclear collisions
before addressing its suppressions. The main uncertainty lies
in the accurate determination of the non-perturbative matrix element for
$J/{\Psi}$ production from the $C\bar C$ pair. Before a decade ago
one used to use the colour singlet model for $J/{\Psi}$ production
at high energy hadronic collisions where it is assumed that the 
$C\bar C$ is produced directly as a colour singlet state which forms
the $J/{\Psi}$ bound state \cite{sing}. 
However, the colour singlet model failed to describe the high momenta 
$J/{\Psi}$ production data at the CDF experiment \cite{cdf}. An effective QCD formalism
based on non-relativistic quantum chromodynamics was developed where
one assumes that the $C \bar C$ is formed as a colour octet state
which finally emits a soft
gluon to form a colour singlet $J/{\Psi}$ \cite{octet}. 
This model (known as colour
octet model) is very successful in explaining the $J/{\Psi}$ production
data at the CDF experiment at Fermi Lab \cite{cdf}. 
In this model the non-perturbative matrix element for 
$J/{\Psi}$ formation from $C \bar C$ is determined from the experiment. 
However, it is not at all clear how to obtain this non-perturbative
matrix element for high energy nucleus-nucleus collisions. The 
complication arises due to the presence of a nuclear/quark-gluon medium.
The second uncertainty lies in the accurate determination of the nuclear
partonic structure function used to compute $C \bar C$. Once these
two issues are solved then one can compute $J/{\Psi}$
production in high energy heavy-ion collisions. 
We will address here how significantly the
heavy quark production cross section
is changed when modified nuclear partonic structure functions
are taken into account. This will have a direct impact on the 
quarkonium production cross section in high energy
heavy-ion collisions at RHIC and LHC.

Before proceeding to our calculation
we briefly mention some of the other effects which must be considered
for a correct calculation of $J/{\Psi}$ suppression.
First of all we mention that the space-time evolution
of quark-gluon plasma in ultra relativistic heavy-ion collisions
is described in various stages, namely; 1) Pre-equilibrium, 2) Equilibrium
and 3) Hadronization.
Once $C\bar C$ is formed, it travels in pre-equilibrium stage of the
quark-gluon plasma. The charm quarks scatter with the light
quarks and gluons in the medium and hence change the energy, momentum
and separation of the $C \bar C$ pairs.
It is not easy to keep track of all the interactions
a charm quark encounters during its propagation in the non-equilibrium
medium. This is because a correct description of non-equilibrium space-time
evolution of the quarks and gluons formed in nuclear collisions is 
not solved accurately. On the other hand, if a $J/{\Psi}$ is formed,
then its interaction with quarks and gluons 
in non-equilibrium medium has also to be taken into account. 
In the equilibrium stage one can argue that they are screened. But a real
dynamical screening argument has to be given. The screening argument given
in \cite{satzs} is valid for a static plasma as calculations were
done in lattice QCD which assumes a static temperature. In heavy-ion 
collisions the temperature is local and space-time dependent and a
screening calculation involving local space-time dependent temperature
is not available. The $J/{\Psi}$ also interacts with the 
hadrons in the hadronic phase \cite{hufner,nayak-jhep}. Hence all these effects 
have to be considered in the study of $J/{\Psi}$ suppression at RHIC and LHC.

In this paper we address one of the above issues, {\it i.e.} we compute
heavy-quark production at RHIC and LHC by taking modifications of the
nuclear partonic distribution functions into account.
The modification of nuclear parton distribution functions
({\it i.e,} $F_2^A(x,Q^2) \neq A~F_2^N(x,Q^2)$ 
and $G^A(x,Q^2) \neq A~G^N(x,Q^2)$)
has been observed in various experiments \cite{nmc-emc}.
Here $F_2^{A(N)}(x,Q^2)$ and $G^{A(N)}(x,Q^2)$ are the quark
and gluon structure functions of the nucleus (nucleon),
$x$ is the momentum fraction of the parton, and $Q^2$ is the momentum
transfer in deep inelastic lepton nucleus (nucleon) scattering.
We take the modified parton distribution functions 
from a previous study \cite{hammon}, where it has been shown that
there is a strong shadowing of the parton distribution functions at 
low $x$. The idea of parton fusion
was first formulated in \cite{qui} and later proven in \cite{qui1}
to appear when the total transverse size 1/Q of the partons in 
a nucleon becomes larger than the proton radius to yield a 
transverse overlapping within a unit of rapidity $xG(x) \ge Q^2R^2$.
The fusion correction in the free nucleon turns out to be significant
only for unusually small values of $x$ and $Q^2$. However, in the case of
heavy nuclei of A $\sim$ 200, the parton recombination is strongly increased
(see \cite{hammon,esk,hammon1}).
This strong shadowing is expected to reduce the heavy quark 
production significantly. We incorporate this feature of gluon shadowing
in the calculation of $C \bar C$ production at RHIC and LHC heavy-ion
collisions.

We use perturbative QCD methods to compute charm and bottom production 
at RHIC and LHC nuclear collisions. 
As we are interested in comparing the results of heavy-quark production
using unshadowed and shadowed parton distributions we will not do a
next-to-leading order calculation for $Q \bar Q$ production. For our
purpose we consider the leading order gluon fusion process 
$g(p_1)+g(p_2) \rightarrow Q(p_3)+\bar{Q}(p_4)$
for heavy quark-antiquark pair production. 
For simplicity a K factor $\sim$ 2 can be multiplied
with this leading order cross section to consider higher
order processes \cite{k-fac}. The gluon fusion process is dominant
at high energy hadron-hadron collisions \cite{john}. 
The squared amplitude for the
above leading order process is given by \cite{combridge}: 

\begin{eqnarray}
\sum |{\mathcal{M}}_{gg \rightarrow Q\bar{Q}}|^2 =
\pi^2\alpha^2(Q^2) [\frac{12}{s^2}(M^2-t)(M^2-u) \nonumber \\
+\frac{8}{3}\frac{(M^2-t)(M^2-u)-2M^2(M^2+t)}{(M^2-t)^2} 
+\frac{8}{3}\frac{(M^2-t)(M^2-u)-2M^2(M^2+u)}{(M^2-u)^2} \nonumber \\
                                 -\frac{2M^2(s-4M^2)}{3(M^2-t)(M^2-u)}
                              -6\frac{(M^2-t)(M^2-u)+M^2(u-t)}{s(M^2-t)}-
6\frac{(M^2-t)(M^2-u)+M^2(t-u)}{s(M^2-u)}],
\end{eqnarray}
where M is the mass of the heavy quark $Q$. The Mandelstam variables $\hat{s}$,
$\hat{t}$, $\hat{u}$ are defined by
\begin{displaymath}
\hat{s}=(p_1+p_2)^2, \hat{t}=(p_1-p_3)^2, \hat{u}=(p_1-p_4)^2
\end{displaymath}
with the four-momenta $p_1$, $p_2$ and $p_3$, $p_4$ of the incident and
the scattered particles respectively. Using the amplitude for the $2 \rightarrow 2$
process, the partonic cross section can be written as
\begin{equation}
\hat{\sigma}(\hat{s}) = \int dt \frac{d\sigma}{dt} =
\frac{1}{16\pi
s^2}\int_{M^2-s(1+\sqrt{1-4M^2/s)}/2}^{M^2-s(1-\sqrt{1-4M^2/s)}/2} dt
\sum\left|{\mathcal{M}}\right|^2.
\end{equation}
Assuming $Q^2=\hat{s}$ \cite{combridge}, the $\alpha(Q^2)$ can be taken
out of the $\hat{t}$ integration to find
\begin{equation}
\hat{\sigma}_{gg\rightarrow{Q\bar{Q}}}(\hat{s})
=\frac{\pi\alpha^2(\hat{s})}{3\hat{s}}
\left[-\left(7+\frac{31M^2}{\hat{s}}\right)\frac{1}{4}\chi+\left(1
+\frac{4M^2}{\hat{s}}
+\frac{M^4}{\hat{s}^2}\right)\log\frac{1+\chi}{1-\chi}\right]
\end{equation}
with
\begin{displaymath}
\chi = \sqrt{1-\frac{4M^2}{\hat{s}}}.
\end{displaymath}
This cross section in the partonic level can be used to compute
the $Q \bar Q$ production cross section in hadronic collisions
by using the parton distribution functions inside the hadron. 
The total cross section for $Q\bar Q$ production
per nucleon in AA collision is:
\begin{equation}
\label{g}
\sigma(s)_{Q\bar{Q}} = \int_{4M^2/s}^1 dx_1 \int_{4M^2/(sx_1)}^1  dx_2 
{g^{A}_g (x_1, Q^2)g^{A}_g(x_2, Q^2)}
\hat{\sigma}_{gg\rightarrow Q\bar{Q}}(\hat{s}),
\end{equation}
where $\hat{s}=x_1x_2 s$, and $g^A(x,Q^2)$ is the gluon structure
function of a bound nucleon inside the nucleus $A$.
On these equations nuclear shadowing effects in
$g^A(x,Q^2)$ will have an influence, as the nuclear
structure functions are reduced in comparison to free nucleon structure
functions at low $x$. We use the shadowing version of parton distribution 
functions in a bound nucleon from ref. \cite{hammon} (with A = 208) together with the
GRV98 set of gluon distribution functions for a free nucleon \cite{grv98}.
The strong coupling constant $\alpha(\hat{s})$ used in the above equation
is also taken from \cite{grv98}. 

We mention here that our calculation is based on primary hard collisions
of partons. These primary hard collisions also produce 
a huge number of
jets and minijets at RHIC and LHC energies. Unlike in pp collisions these
minijets suffer from secondary collisions because they are large in 
number. Secondary collisions among these primarily produced
minijets bring the system towards equilibrium. The secondary collisions 
among these minijets can also produce additional $C \bar C$ pairs.
However, it should be checked whether this heavy quark production
is small or large due to their large masses.
It can be argued that the secondary charm production is very small
for an equilibrium quark-gluon plasma because the production rate behaves
like $\sim ~e^{-M/T}$. However, in the early pre-equilibrium
stage of the quark-gluon plasma, 
the average energy carried by a parton is large
and hence their secondary collisions can produce significant secondary
$C\bar C$ pairs. One needs a real calculation by using 
non-equilibrium distribution of the minijet plasma at RHIC and LHC
to see whether the secondary $C\bar C$ pair production becomes larger/smaller
than the primary charm production. 
The secondary charm production calculation is very important
because a huge production of secondary charm may suggest an enhancement
of $J/{\Psi}$ at RHIC and LHC rather than its suppression. 
In any case, the shadowing will also
decrease the minijet production \cite{hammon}. In such situations
the minijet temperature will also be reduced which in turn will decrease
secondary
charm production. Hence as far as shadowing of initial parton distributions
is concerned, it decreases both primary and secondary charm productions.
In the following we will present results of the cross section for 
heavy quark-antiquark pair production from primary partonic collisions
at different center of mass energies to observe the effect of shadowing
on quarkonium production.

In Fig.1 we present the cross section (in mb)
for heavy quark/anti-quark pair production
at RHIC energy as a function of the heavy-quark mass. The dotted line is
obtained by using no-shadowing gluon distribution functions, {\it i.e,} 
 we have used $g^A(x,Q^2)=g^N(x,Q^2)$, with $g^N(x,Q^2)$ taken from
the GRV98 parametrization \cite{grv98} (Note that $g^A(x,Q^2)$ used
in Eq. (\ref{g}) is the distribution function of a bound nucleon inside
a nucleus A). It can be mentioned that  
our results are with respect to the nucleon-nucleon scattering in nuclear
collisions. To estimate the total cross section in nucleus-nucleus
collisions, 
one 
has to multiply our
results by $A^2$. The solid line is obtained by using strong shadowing
of gluons at high energy \cite{hammon}. It can be seen that the shadowing
in gluon distribution function decreases the heavy quark production cross
section significantly. The reduction of charm
quark production cross section at RHIC and LHC is interesting
because the suppression of
$J/{\Psi}$ may be due to the shadowing of the initial gluon distribution
functions inside the nucleus rather than existence of QGP. 
However, before making any conclusions,
one has to compute $J/{\Psi}$ production by using a suitable model
for hadronization. In addition to this,
one has to compute secondary charm quarks which are
produced from the rescattering of minijets being
formed in high energy heavy-ion collisions.
We have plotted the ratio of heavy-quark production cross sections
with and without shadowing in the same figure. This is
represented by the dashed line. Our results cover both charm and
bottom quark production at RHIC. In future RHIC will provide an opportunity 
to measure open charm production.

In Fig.2 we present the cross section for heavy quark/anti-quark
pair production at LHC energy as a function of the heavy-quark mass. 
We present our results up to $M_Q \sim$ 200 GeV which covers
charm, bottom and top quark production at LHC. Dotted, dashed and solid
lines correspond to the same situation as in Fig.1.
The difference in charm and bottom quark production at LHC
is found to be much larger than that at RHIC. It can be seen from
the figure that the shadowing effect on charm production is much larger
than that on bottom production. The ratio is close to 1 for the 
production of the top quark.  

In Fig.3 we present the $C\bar C$ production cross section
as a function of $\sqrt{s}$. The dotted line is for the case without
shadowing, 
and the solid line is with shadowing of gluons. It can be seen that 
the $C\bar C$ production
cross section at RHIC for both cases is not much different. 
However, the difference becomes quite large when the calculation is
repeated for very high
energy nuclear collisions. At LHC energy the reduction in the cross section
is quite large, about a factor of 10. The same is true for $B\bar B$
production at different energies, see Fig.4. We mention here that 
our results are obtained from the leading order partonic scattering
processes. Our results can be multiplied by a K factor ($\sim$ 2) to
take higher order processes into account. 

To summarize, we have computed heavy quark/anti-quark pair production
cross sections per nucleon-nucleon collision 
in RHIC and LHC heavy-ion collisions with shadowing of parton
structure functions taken into account. 
It is shown that the shadowing of parton distribution functions 
significantly
decreases the production cross section of heavy quarks at LHC. 
At RHIC energy this difference is not very large but still 
significant. One can obtain the total charm 
production at RHIC and LHC from these cross sections via the relation
$N = T(0) \sigma$, where the Glauber geometrical
factor $T(0) \sim ~\frac{A^2}{\pi R_A^2}$ for the impact parameter b=0 fm. 
This geometrical factor represents the total number of
binary nucleon-nucleon collisions in central AA collisions,
which behaves like $A^{4/3}$ \cite{eskaj}. Although the difference of
the $b\bar{b}$ and $c\bar{c}$ cross sections for shadowing and 
no shadowing at RHIC is not large,
there is a significant difference in the total number of charm
and bottom quarks for the two versions of shadowing. This has a direct
impact on quarkonium and open charm production at RHIC. 
We suppose that a reduction of the charm quark cross section due to
shadowing is very significant because $J/{\Psi}$ suppression 
might occur due to the depletion of the initial
gluon distribution functions rather than due to the existence 
of a quark-gluon plasma or due to the presence of hadronic medium.

\acknowledgements

We thank  Lars Gerland, Nils Hammon and Chung-Wen Kao for discussions. 
G. C. N. acknowledges the financial
support from Alexander von Humboldt Foundation.

\newpage

\section*{Figure Captions}

\textbf{Fig. 1.} The cross section for heavy-quark production
via the parton fusion
process $gg \rightarrow Q \bar Q$ at RHIC. The solid (dotted) line represents
the heavy-quark production cross section (in $mb$) with (without) gluon
shadowing (explained in the text). The dashed line represents the ratio 
of the cross sections with and without shadowing.  \\
  
\textbf{Fig. 2.} The cross section for the heavy-quark production
via the parton fusion
process $gg \rightarrow Q \bar Q$ at LHC. The solid (dotted) line represents
the heavy-quark production cross section (in $mb$) with (without) gluon
shadowing (explained in the text). The dashed line represents the ratio 
of the cross sections with and without shadowing.  It can be seen that 
there is a strong shadowing effect on charm production (about a factor
of 10, (see the ratio curve and discussion in the text)). \\

\textbf{Fig. 3.} The cross section in $mb$ from the process
$gg \rightarrow C \bar C$. Solid (dotted) line is
with (without) gluon shadowing (explained in the text). 
$M_C=1.5~GeV$ is used in the calculation. \\

\textbf{Fig. 4.} The cross section in $mb$ from the process
$gg \rightarrow B \bar B$. Solid (dotted) line is
with (without) gluon shadowing (explained in the text).
$M_B=5.4~GeV$ is used in the calculation.

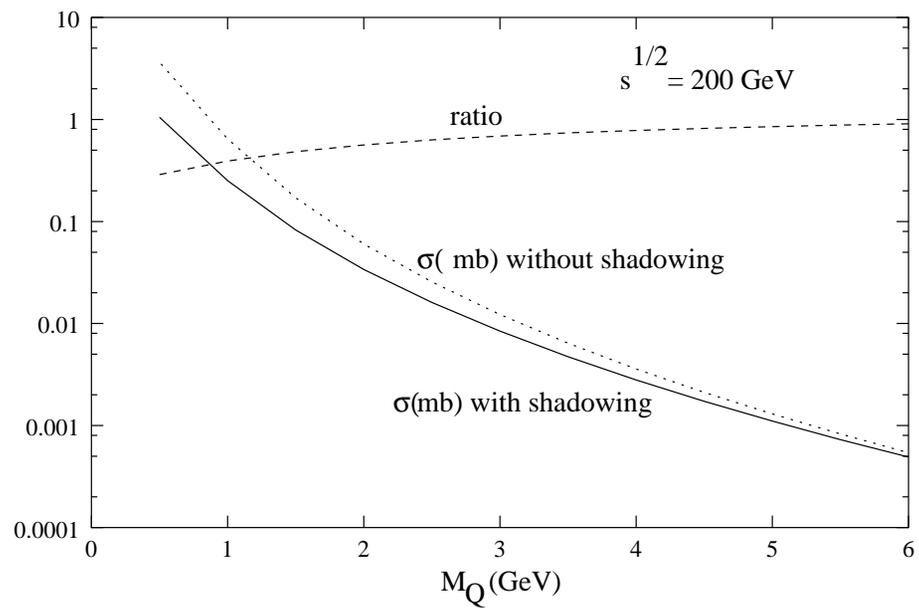
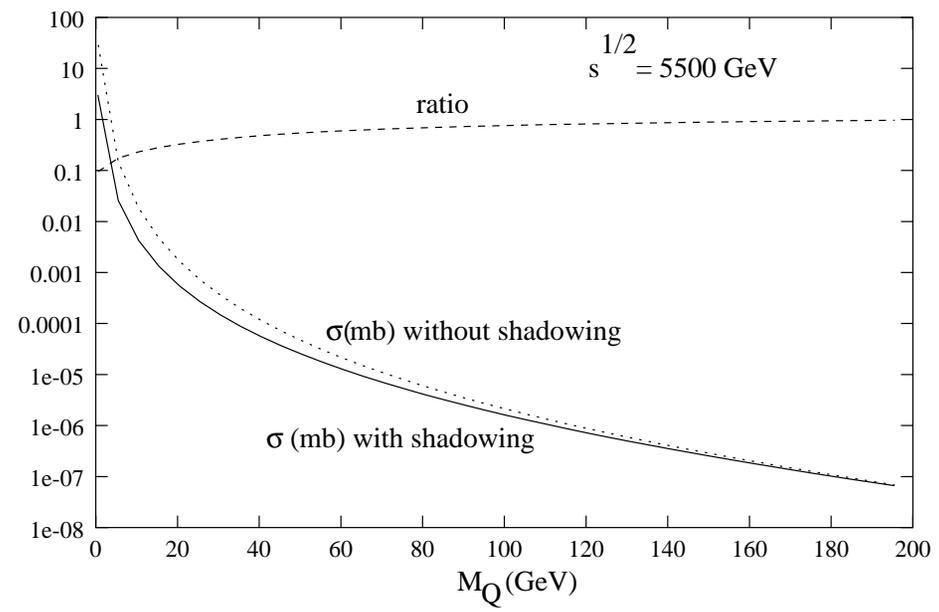
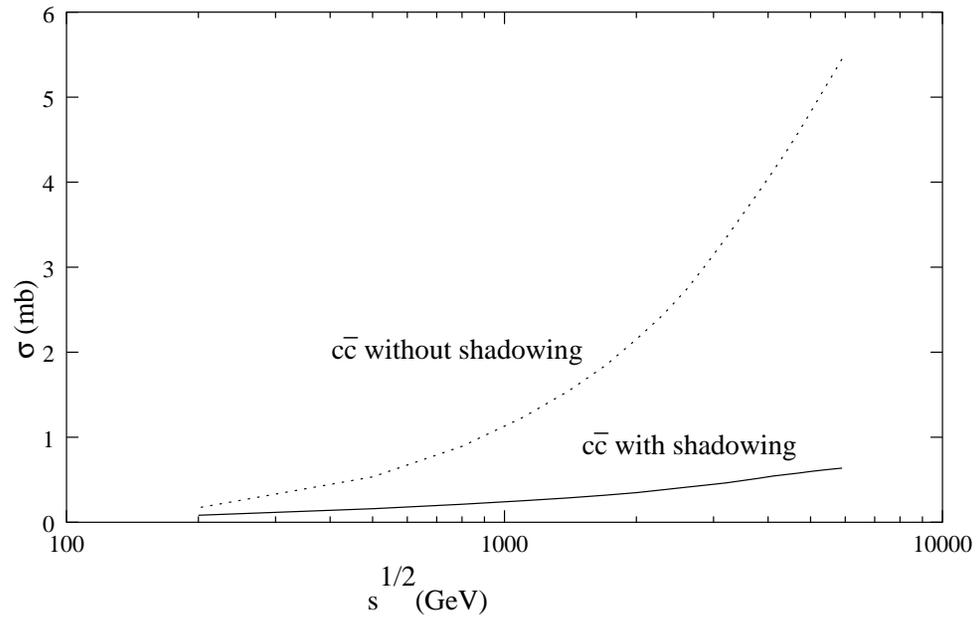
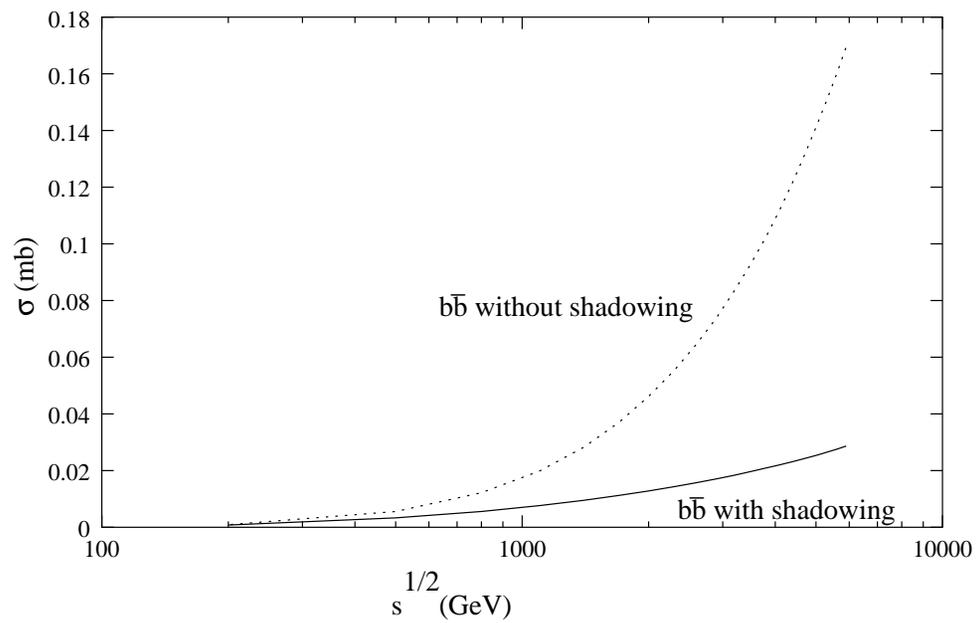
\end{document}